\documentclass[twocolumn]{emulateapj}
\usepackage{graphicx}
\usepackage{epsfig}
\usepackage{dcolumn}
\usepackage{rotating}
\shorttitle{Short-term variability of 3C 390.3}
\shortauthors{Gliozzi et al.}
\def\3c{3C~390.3}

\def\feka{Fe K$\alpha$}
\def\chandra{{\it Chandra}} 
\def\xmm{{\it XMM-Newton}}
\def\suzaku{{\it Suzaku}}
\def\asca{{\it ASCA}} 
 
\def\rxte{{\it RXTE}} 
\def\sax{{\it BeppoSAX}} 
 
\def\rosat{{\it ROSAT}}

\def\lum{erg s$^{-1}$}
\def\flux{erg cm$^{-2}$ s$^{-1}$}

\def\lsim{\ifmmode{\;\mathop{}^{<}_{\sim}\;}\else{$\;\mathop{}^{<}_{\sim}\;$}\fi}
\def\gsim{\ifmmode{\;\mathop{}^{>}_{\sim}\;}\else{$\;\mathop{}^{>}_{\sim}\;$}\fi}
\def\ltsima{$\; \buildrel < \over \sim \;$}
\def\simlt{\lower.5ex\hbox{\ltsima}} 
\def\gtsima{$\; \buildrel > \over \sim \;$}
\def\simgt{\lower.5ex\hbox{\gtsima}} 

\begin{document}
\title{Short-term variability and PSD analysis of the radio-loud AGN 3C 390.3}

\author{Mario Gliozzi}
\affil{George Mason University, 4400 University Drive, Fairfax, VA 22030}

\author{Iossif E. Papadakis}
\affil{Physics Department, University of Crete, Greece}

\author{Michael Eracleous\altaffilmark{1}}
\affil{Department of Astronomy \& Astrophysics, The Pennsylvania State University, 525 Davey Lab, University Park, PA 16802}
\altaffiltext{1}{Center for Gravitational Wave Physics, The Pennsylvania State University, University Park, PA 16802}

\author{Rita M. Sambruna}
\affil{NASA's Goddard Space Flight Center, Code 661, Greenbelt, MD 20771}

\author{David R. Ballantyne}
\affil{Department of Physics, The University of Arizona, 1118 East 4th Street, Tucson, AZ 85721}

\author{Valentina Braito}
\affil{Department of Physics and Astronomy, University of Leicester, Leicester LE1 7RH, UK}

\author{James N. Reeves}
\affil{Astrophysics Group, School of Physical and Geographical Sciences, Keele University, Keele, Staffordshire, UK}

\begin{abstract}
We investigate the short-term variability properties and the power spectral
density (PSD) of the  Broad-Line Radio Galaxy (BLRG) 3C 390.3 using
observations made by \xmm, \rxte, and \suzaku\ on several occasions
between October 2004 and December 2006. The main aim of this work is to
derive model-independent constraints on the origin of the X-ray emission and 
on the nature of the central engine in \3c. 
On timescales of the order of few hours, probed by uninterrupted \xmm\
light curves, the flux of
\3c\ is consistent with being constant in all energy bands. On
longer timescales, probed by the 2-day \rxte\ and \suzaku\
observations, the flux variability becomes significant.  The latter
observation confirms that the spectral variability behavior of \3c\ is
consistent with the spectral evolution observed in (radio-quiet)
Seyfert galaxies: the spectrum softens as the source brightens.
The correlated variability between soft and hard X-rays, observed
during the \suzaku\ exposure and
between the 2 \xmm\ pointings, taken 1 week apart, argues against
scenarios characterized by the presence of two distinct variable
components in the 0.5--10 keV X-ray band.
A detailed PSD analysis carried out over five decades in frequency
suggests the presence of a break at $T_{br}=43^{+34}_{-25}$ days at a
92\% confidence level.  This is the second tentative detection of a PSD
break in a radio-loud, non-jet dominated AGN, after the BLRG 3C~120, 
and appears to be in
general agreement with the relation between $T_{br}$, $M_{\rm BH}$,
and $L_{\rm bol}$, followed by Seyfert galaxies.  Our results indicate
that the X-ray variability properties of \3c\ are broadly consistent
with those of radio-quiet AGN, suggesting that the X-ray emission
mechanism in \3c\ is similar to that of nearby Seyfert galaxies without
any significant contribution from a jet component.

\end{abstract}

\keywords{Galaxies: active -- Galaxies: nuclei -- X-rays: galaxies}

\section{Introduction}

Relativistic bipolar outflows, generally emitting most of their energy
in the radio range, are one of the most dramatic manifestations of the
presence of supermassive black holes in Active Galactic Nuclei (AGN).
Although it is widely accepted that these ejections are closely
related to the accretion process onto the central black hole, the
details of this link are still unknown.

A promising approach for tackling this problem is to investigate the
X-ray properties of AGN with jets (generally called radio-loud AGN)
and carry out a systematic comparison with their radio-quiet
counterparts, the Seyfert galaxies. Unlike optical and UV light,
X-rays are not significantly attenuated and are less affected by
dilution from the host galaxy  and are thought to be produced
in the inner most regions of the accretion flow, thus may provide 
the most direct view of the central engine.

Broad-line radio galaxies (hereafter BLRGs) are one of the best
classes of radio-loud AGN for this comparative analysis; they have
optical and UV spectral properties similar to Seyfert galaxies, but
they also host large-scale radio jets that are absent in their
radio-quiet counterparts.  Past X-ray spectroscopic studies, employing
\asca, \rxte\, and \sax\ data, have shown that BLRGs have weak \feka\ lines
and weak or absent Compton reflection humps at energies $\gsim
10$~keV, a hallmark of Seyfert 1 galaxies (e.g., Wo\'zniak et
al. 1998; Sambruna et al. 1999; Eracleous et al. 2000; Zdziarski \&
Grandi 2001; Grandi et al. 2006). These findings have been confirmed
by recent studies that made use of higher quality spectra provided by
\chandra\ and \xmm\ (e.g., Ballantyne et al. 2004, Ballantyne 2005;
Ogle et al. 2004; Lewis et al. 2005; Gliozzi et al. 2007). Indeed, the
weakness of the \feka\ line and the Compton reflection continuum are
very important observational clues, since they represent a major
difference between radio-loud and radio-quiet AGN. However, the origin
of this difference is still debated (see Gliozzi et al. 2007 for a
detailed discussion on the competing models).

The importance of temporal studies lies in that they may provide
model-independent information that complements the findings from
spectral studies and possibly breaks the spectral degeneracy.  Indeed,
the similarity of the temporal and spectral variability properties of
2 BLRGs (including \3c) with those of Seyfert galaxies, led us to rule
out a jet origin for the bulk of X-ray flux from these BLRGs and
provided tight constraints on the jet contribution (Gliozzi et
al. 2003a). 

Past temporal studies indicate that the flux variability of \3c\ is associated
with spectral variability, in the sense that the spectrum softens as the
flux increases. The presence of flux and spectral variability has been
observed separately in the 2--15 keV band with \rxte\ (Gliozzi et al. 2003a,
2006), in the 2-10 keV energy band using \asca\ and  {\it Ginga} 
(Leighly et al. 1997), as well as at softer energies (0.1--2.4 keV)
with \rosat\ (Leighly et al. 1997).
Importantly, all the above results, are based either on long-term (months
to years) monitoring campaigns or on multiple observations spanning several
years, but none addresses specifically the short-term variability. Although
in the literature there are
several studies based on individual observations of \3c\ with different X-ray
satellites (e.g., {\it EXOSAT} from Inda et al. 1994; \asca\ from Eracleous 
et al. 1996, or \sax\ from Grandi et al. 1999), they all are focused on 
the spectral analysis and the temporal analysis is generally limited to few
sentences indicating that the flux appears to be constant on timescales
shorter than 1 day. 

The apparent absence of short-term variability in \3c\
seems to be in line with scaling relations inferred from power spectral 
density studies of radio-quiet AGN (see McHardy et al. 2006 and references
therein). However, the lack of short-term variability in this AGN has never 
been tested by a satellite with the capabilities of \xmm, which combines a 
very high throughput with highly elliptical orbits. These capabilities
produce high quality uninterrupted light curves that have revealed the
presence of short-term variability also in unexpected AGN classes, such as
LINERs or low-luminosity Seyfert galaxies (e.g., Gliozzi et al. 2003b, 2008;
Papadakis et al. 2009).

Taking advantage of the unique capabilities of \xmm\ we perform for the first 
time a thorough analysis of the temporal and spectral variability of \3c\ on
timescales of a few hours. This study is complemented by a similar analysis
on timescales of 2 days, based on high quality variability data from \suzaku\ 
and from \rxte. These data are then combined with long-term \rxte\ monitoring
data to produce the first power spectral density (PSD) of \3c.
A detailed analysis of the time-averaged spectral properties is reported
in a companion paper by Sambruna et al. (2009) and can be summarized
as follows: 1) the broad-band 0.4--100 keV continuum is well described
by a power law with $\Gamma=1.6$ and a high-energy cut-off at $E_{\rm
cutoff}=$175 keV; 2) reprocessing by two different ``reflectors'' (one
neutral with $R=0.5$ and one ionized with $\xi\simeq$2700) is required;
3) the \feka\ line profile is well fitted by a narrow component
centered at 6.4 keV plus a broad component at 6.6 keV (apparently from
He-like Fe).

This paper is organized as follows. In $\S~2$ we describe the
observations and data reduction. The short timescale (from few hours to
2 days) flux and
spectral variability analyses are reported in $\S~3$ and $\S~4$,
respectively. In $\S~5$ we perform a PSD analysis combining the
short-time scale light curves from this paper with long-time scale
light curves from past \rxte\ monitoring campaigns.  In $\S~6$ we
discuss the main results, and finally in $\S~7$ we summarize the main
conclusions.  Hereafter, we adopt $H_0=71{\rm~km~s^{-1}~Mpc^{-1}}$,
$\Omega_\Lambda=0.73$ and $\Omega_{\rm M}=0.27$ (Bennet et
al. 2003); with the assumed cosmological parameters, the luminosity
distance of \3c\ (z=0.056) is 247 Mpc.
For the temporal analysis we make
use of the $\chi^2$ test to assess the significance of the variability
and consider a light curve significantly variable if the probability
of the null hypothesis (i.e., the source being constant) is less than
1\%.

\section{Observations and Data Reduction}

We observed \3c\ with \xmm\ on 2004 October 10 and 17 for 50 ks and 20
ks, respectively.  All of the EPIC cameras (Str\"uder et al. 2001;
Turner et al. 2001) were operated in small window mode to prevent
photon pile-up, and with medium filters, due to the presence of bright
nearby sources in the field of view. The data reduction has been
performed following the standard procedure with 
the \xmm\ Science Analysis Software (\verb+SAS+) 7.1; a detailed 
description is given in Sambruna et al. (2009).
\subsection{XMM-Newton}
\begin{figure*}
\includegraphics[bb=80 3 440 410,clip=,angle=0,width=8.cm]{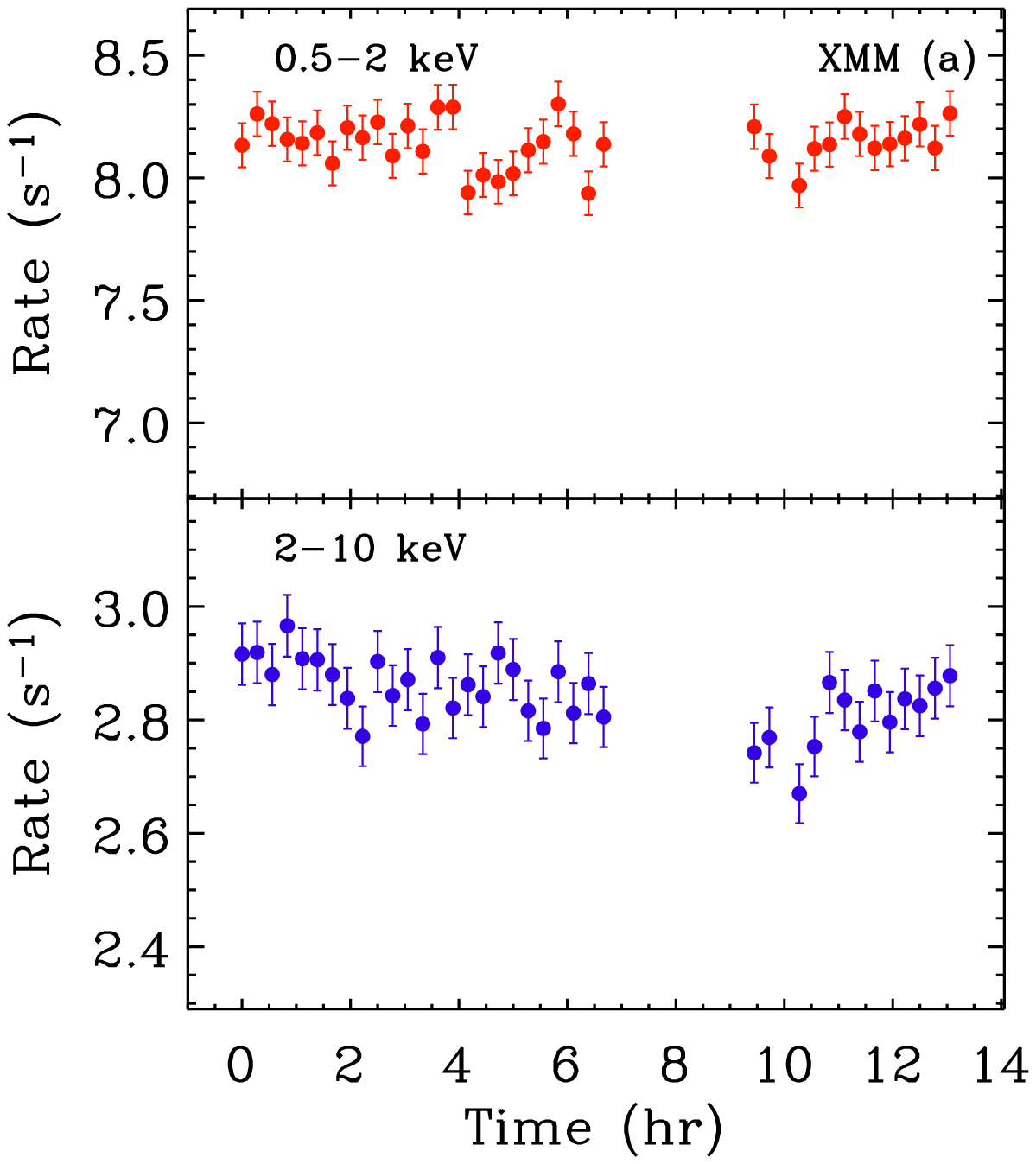}\includegraphics[bb=80 3 440 410,clip=,angle=0,width=8.cm]{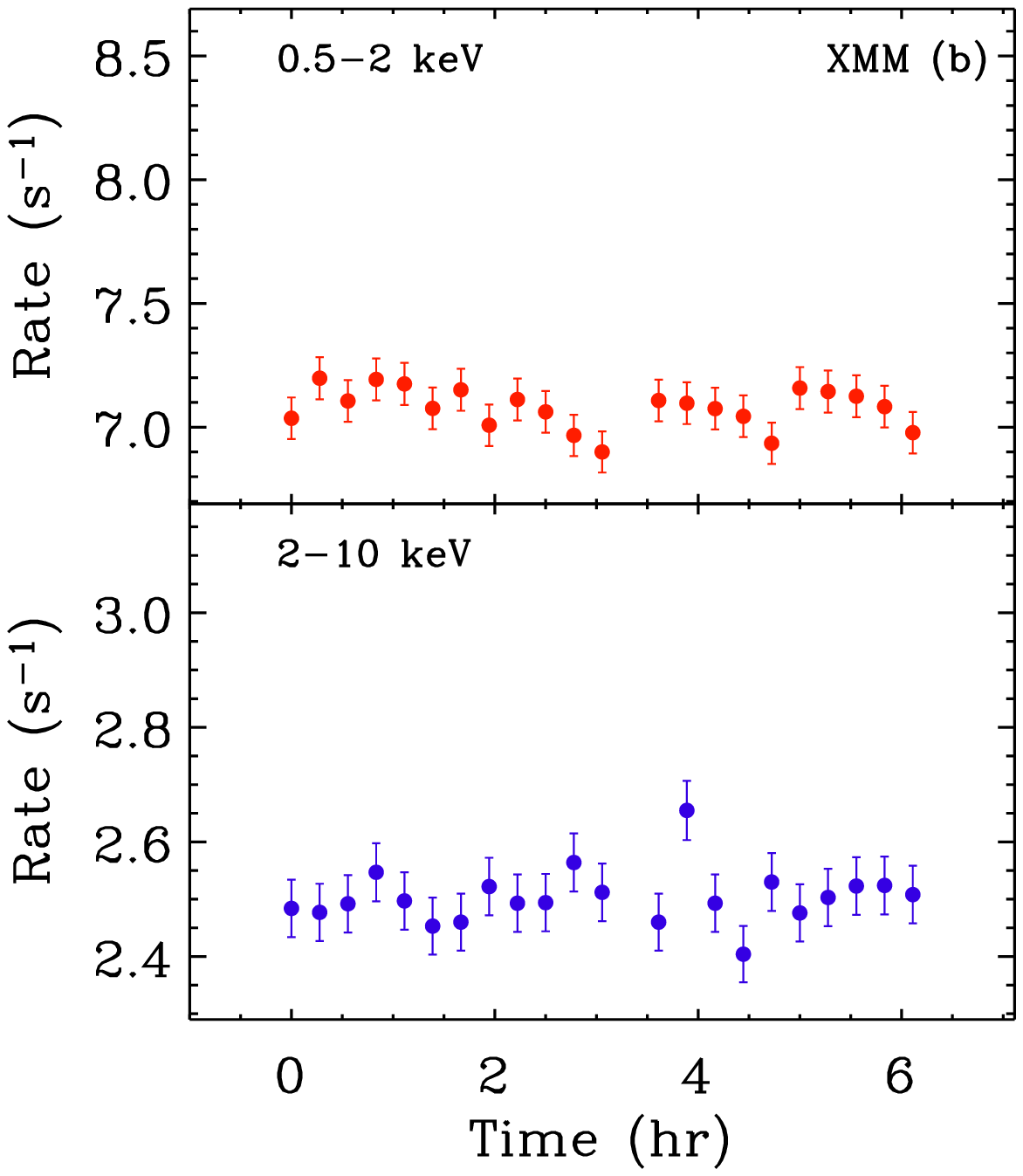} 
\caption{{\it Left:} \xmm\ EPIC pn light curves of the soft 
(0.5--2 keV) and hard count rate (2--10 keV) on 2004 October 10. 
Time bins of 1000 s have been used.
{\it Right:} \xmm\ EPIC light curves light curves on 2004 October 17.
} 
\label{figure:fig1} 
\end{figure*}

In order to cover the harder part of the X-ray band, up to 40~keV,
\3c\ was also observed by \rxte. Unfortunately, the \rxte\ coverage,
which was intended to be simultaneous to the \xmm\ observations, was
carried out on 2005 January 12 and 13, i.e. nearly two months after
the \xmm\ observations, due to scheduling problems.  Because of the
well known long-term temporal and spectral variability of \3c, the
\rxte\ data cannot be safely combined with the EPIC data for a
broadband spectral analysis. Nevertheless, the \rxte\ observation,
which has a total exposure of 80 ks spanning a 2-day interval, is
useful for investigating the variability on intermediate timescales at
higher energies.

The \rxte\ observations (both the new observations presented here and
the older observations used for the PSD analysis) were carried out
with the Proportional Counter Array (PCA; Jahoda et al. 1996), and the
High-Energy X-Ray Timing Experiment (HEXTE; Rotschild et al. 1998)
instruments. Here we will consider only PCA data, because the
signal-to-noise ratio (hereafter S/N) of the HEXTE data is too low for
a meaningful analysis. The PCA data were screened according to the
following acceptance criteria: the satellite was out of the South
Atlantic Anomaly (SAA) for at least 30 minutes, the Earth elevation
angle was $\geq 10^{\circ}$, the offset from the nominal optical
position was $\leq 0^{\circ}\!\!.02$, and the parameter ELECTRON-2 was
$\leq 0.1$. The last criterion excludes data with high particle
background rates in the Proportional Counter Units (PCUs). The PCA
background light curves were determined using the ${\rm L}7-240$ model
developed at the \rxte\ Guest Observer Facility (GOF) This model is
implemented by the program {\tt pcabackest} v.2.1b and is applicable
to ``faint'' sources, i.e., those with count rates $< 40 {\rm
s^{-1}~PCU^{-1}}$. All the above tasks were carried out with the help
of the \verb+REX+ script provided by the \rxte\ GOF, which calls the
relevant programs from the {\tt FTOOLS} v.6.5 software package and
also produces response matrices and effective area curves for the
specific time of the observation. Data were initially extracted with
16~s time resolution and then re-binned to different bin widths for
different applications.  The short-term temporal analysis is
restricted to PCA, STANDARD-2 mode, 2--15 keV, Layer 1 data, because
that is where the PCA is best calibrated and most sensitive. For the
PSD study we restricted the analysis to the 2--10 keV energy band,
since this is the common energy range for \rxte, \suzaku, and \xmm.
PCUs 0 and 2 were turned on throughout the monitoring
campaign. However, since the propane layer on PCU0 was damaged in May
2000, causing a systematic increase of the background, we
conservatively use only PCU2 for our analysis. All quoted count rates
are therefore for one PCU.

\suzaku\ observed 3C 390.3 on 2006 December 14--16 for a total
exposure time of 100 ks. We used the cleaned event files obtained from
version 2 of the Suzaku pipeline processing, according to standard
screening criteria. The XIS0, XIS1, and XIS3 source light curves were
extracted from circular regions of radius 
2{\farcm}9 centered on the source and combined in order to increase
the S/N; background light curves were extracted from four circular
regions offset from the source.  For the HXD-PIN data reduction and
analysis we followed the latest \suzaku\ data reduction guide, and
used the rev2 data, which include all the 4 cluster units and the best
background available, model D, which has a systematic uncertainty of
$\pm1.3\%$ at 1 $\sigma$
level\noindent\footnote{ftp://legacy.gsfc.nasa.gov/suzaku/doc/hxd/suzakumemo-2008-03.pdf}.
See Sambruna et al. (2008) for a more detailed description of the
\suzaku\ data reduction.

\begin{figure*} 
\includegraphics[bb=90 20 430 460,clip=,angle=0,width=8.cm]{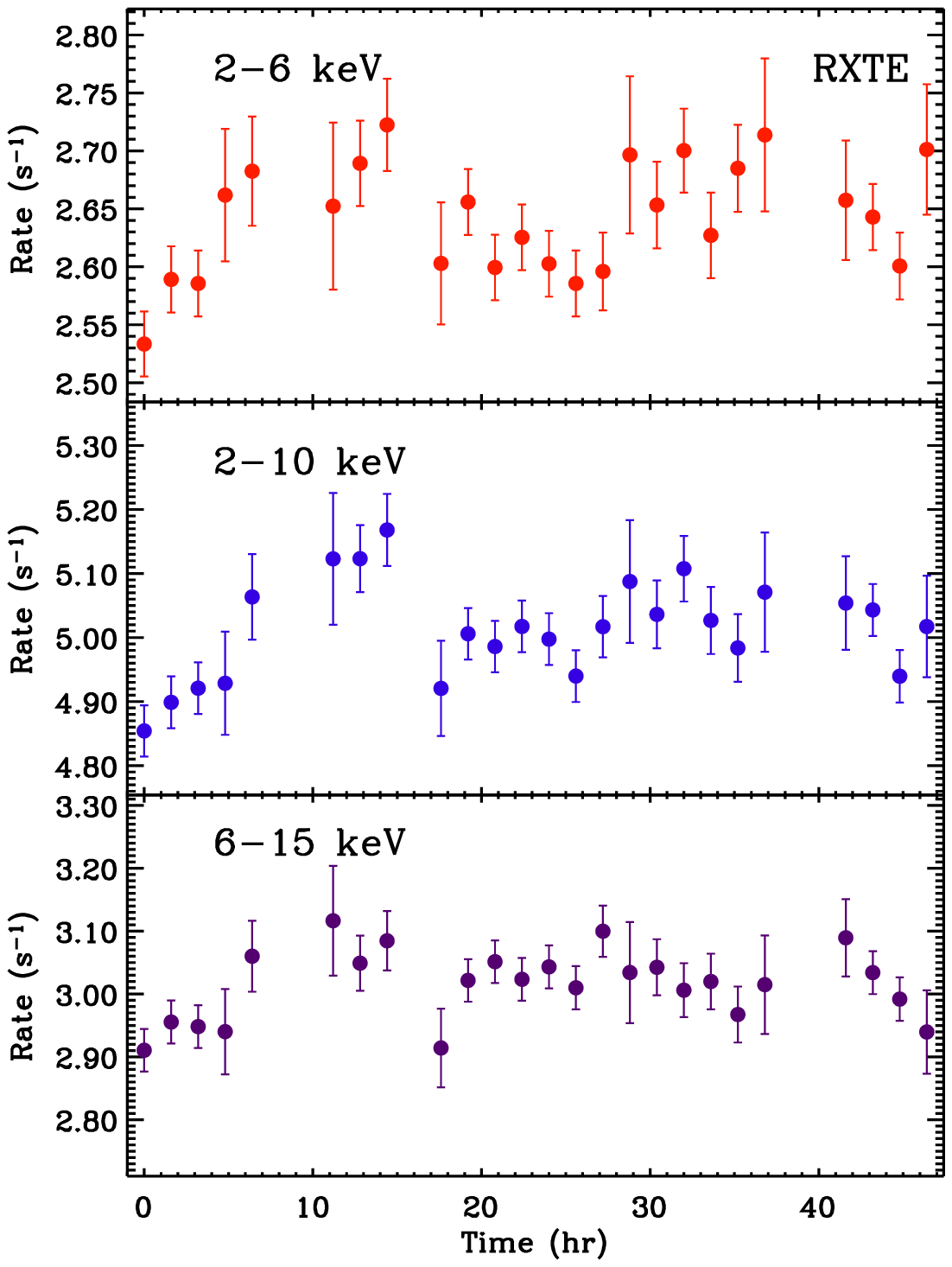}\includegraphics[bb=90 20 430 460,clip=,angle=0,width=8.cm]{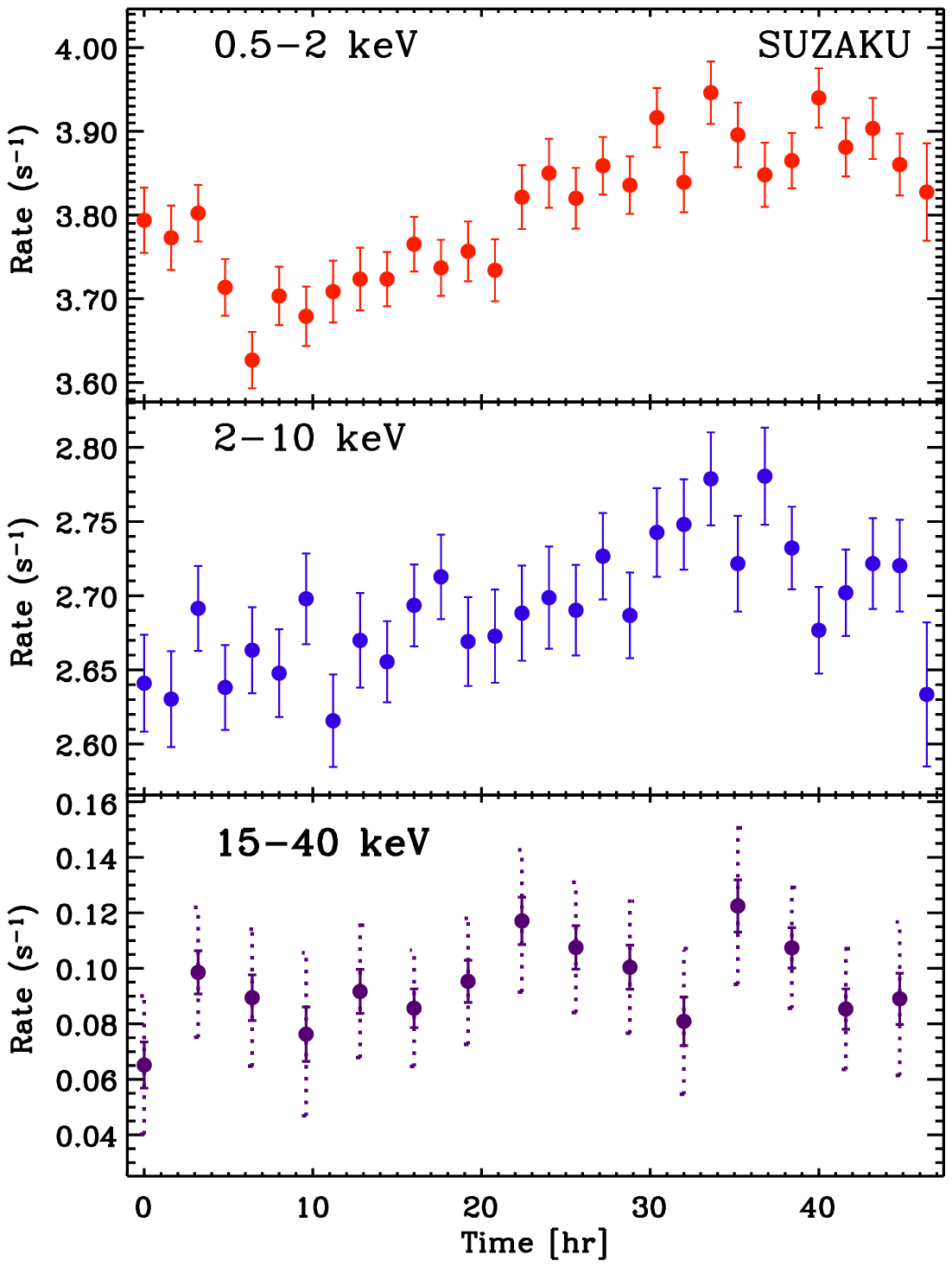} 
\caption{{\it Left:} \rxte\ PCA light curves in the
2--6 keV, 2--10 keV and 6--15 keV energy bands. Time bins of 5760 s 
($\sim$ 1 \rxte\ orbit) have been used.
{\it Right:} \suzaku\ XIS013 light curves in the 
0.5--2 keV and 2--10 keV energy bands (top and middle panels).Time bins of 
5760 s ($\sim$ 1 \suzaku\ orbit) have been used. The bottom panel shows the 
\suzaku\ PIN light curve in the 15--40 keV energy range; time bins are
2 satellites orbit. Given the uncertainty on the PIN background, for 
completeness, we have plotted both 1 (solid line) and 3 $\sigma$ 
(dotted lines) error bars.
} 
\label{figure:fig2} 
\end{figure*}
\section{Short-Time Scale Flux Variability}
 
Figures~\ref{figure:fig1}a and \ref{figure:fig1}b show the EPIC pn
light curves of the soft (0.5--2 keV; top panels) and hard (2--10 keV;
bottom panels) count rate on 2004 October 10 and October 17.
Hereafter, for the sake of simplicity, we will refer to the 2 \xmm\
pointings as observations A and B, respectively. The average 2--10~keV
fluxes during these two observations were $3.9\times10^{-11}$ and
$3.4\times10^{-11}$\flux, respectively.  In order to allow a direct
comparison between observations A and B, we have kept the same
vertical scales in Fig.~\ref{figure:fig1}a and b. In this way, the
significant decrease in count rate is easily discernible.
Specifically, between the first and the second exposure, the EPIC pn
average count rate decreases from $10.99 \pm 0.01~{\rm s}^{-1}$ to
$9.58\pm 0.01~{\rm s}^{-1}$ in the 0.5--10 keV energy band.  A visual
inspection of Fig.~\ref{figure:fig1} suggests that, despite the
presence of some small-amplitude variations during observation A,
within each individual exposure the soft and hard count rates do not
vary significantly.  The lack of short-term variability is formally
confirmed by a $\chi^2$ test
and by the fractional variability analysis, whose results are
reported in Table 1 and Table 2, respectively.

\begin{table*} 
\footnotesize
\caption{Flux variability properties}
\begin{center}
\begin{tabular}{cccccccc} 
\hline        
\hline
\noalign{\smallskip}  
Date &  Instrument (Satellite) & $F_{\rm 2-10 keV}$             &  \multicolumn{2}{c}{0.5--2 keV variability}       &   \multicolumn{2}{c}{2--10 keV variability}  &  \\ 
(yyyy/mm/dd)     &             &   (${\rm erg~cm^{-2}~s^{-1}}$)  &    $\chi^2$/dof    &  $P_\chi^2$ &  $\chi^2$/dof    &  $P_\chi^2$ \\
\hline 
\noalign{\smallskip}
2004/10/10       & EPIC pn (XMM)  & $3.9\times 10^{-11}$                                &    41.08/37        & 0.30        & 47.20/37  & 0.12 \\      
\noalign{\smallskip}
2004/10/17       & EPIC pn (XMM)  & $3.4\times 10^{-11}$                                &    20.01/21        & 0.52        & 19.07/21  & 0.58 \\      
\noalign{\smallskip}
\hline  
\noalign{\smallskip}
 2005/01/12-13    & PCA (RXTE)  & $5.5\times 10^{-11}$                           &      &   &  58.37/24 & $1.1\times 10^{-4}$  \\ 
\noalign{\smallskip}
\hline
\noalign{\smallskip}
 2006/12/14-16    & XIS (Suzaku)  & $3.1\times 10^{-11}$                         &   105.24/29   & $< 1\times 10^{-6}$  &  42.97/29 & $4.6\times 10^{-2}$  \\ 
\noalign{\smallskip}
\hline
\hline
\end{tabular}
\end{center}
\label{tab1}
\footnotesize
\end{table*}   
\begin{table*} 
\footnotesize
\caption{Spectral variability properties}
\begin{center}
\begin{tabular}{cccccccc} 
\hline        
\hline
\noalign{\smallskip}
Date &  Instrument (Satellite) &   \multicolumn{2}{c}{HR variability}      &  $F_{\rm var,soft}^a$      &   $F_{\rm var,hard}^b$   \\ 
(yyyy/mm/dd)     &             &  $\chi^2$/dof    &  $P_\chi^2$ &&                          &                         \\
\hline 
\noalign{\smallskip}
2004/10/10       & EPIC pn (XMM)  &    38.92/37 & 0.38           &    $(3.6\pm4.3)\times10^{-3}$        &  $(9.6\pm5.2)\times10^{-3}$\\ 
\noalign{\smallskip}
2004/10/17       & EPIC pn (XMM)  &    21.61/21 & 0.42           &    $\dots$           &  $\dots$\\
\noalign{\smallskip}
\hline  
\noalign{\smallskip}
 2005/01/12-13    & PCA (RXTE)  &      24.22/24 & 0.45           &    $(9.3\pm5.2)\times10^{-3}$ & $(8.4\pm5.9)\times10^{-3}$\\
\noalign{\smallskip}
\hline
\noalign{\smallskip}
 2006/12/14-16    & XIS (Suzaku)  &    37.7/29 & 0.13            &   $(1.9\pm0.2)\times10^{-2}$ & $(1.1\pm0.3)\times10^{-2}$\\
\noalign{\smallskip}
\hline
\hline
\end{tabular}
\end{center}
$^a$ The soft band is 0.5--2 keV for \xmm\ and \suzaku, whereas for \rxte\ it is 2--6 keV.\\
$^b$ Similarly, the hard band corresponds to 2--10 keV for  \xmm\ and \suzaku, and to 6--15 keV for \rxte.
\label{tab2}
\footnotesize
\end{table*}  
\subsection{RXTE}

\rxte\ observed \3c\ on 2005, January 12 and 13, when the source was
in a very high brightness state: the average 2--10~keV flux was
$5.5\times10^{-11}$\flux\ with a corresponding luminosity of
$4.1\times10^{44}$\lum, which is slightly higher than the maximum
value registered in the two-year \rxte\ monitoring campaign (Gliozzi
et al. 2006).  The left panel of Figure~\ref{figure:fig2} shows the
light curves in the 2--6 keV (top panel; $P_{\chi^2}=1.5\times10^{-3}$), 
2--10 keV (middle panel; $P_{\chi^2}=1.1\times10^{-4}$),
and 6--15 keV (bottom panel; $P_{\chi^2}=4.9\times10^{-2}$) energy bands, 
respectively.  Since the
\rxte\ PCA is best calibrated in the 2--15 keV energy band, a direct
comparison with the soft EPIC pn light curve described above cannot be
performed. Nevertheless, \rxte\ allows a direct comparison in the
2--10 keV range and also the investigation of the harder X-rays up to
15 keV.  On timescales of the order of 2 days, the source count rate
is clearly variable in the 2--10 keV energy band (Table 1). The fact that
the harder band is only marginally variable (it is variable at a 95\% 
confidence level) can be ascribed to the lower data quality at higher energies.

\subsection{Suzaku}

\suzaku\ observed \3c\ on 2006 December 14--16, when the source was in
a brightness state similar to that observed by \xmm\ 2 years earlier
($F_{\rm2-10~keV}=3.1\times10^{-11}$\flux). The XIS 0.5--2 keV and
2--10 keV light curves are shown in the top and middle panels of
Fig.~\ref{figure:fig2}b. Just as with the \rxte\ PCA light curves,
both XIS light curves show significant (and correlated) variability on 
timescales of 2
days, confirmed by a $\chi^2$ test whose results are reported in Table
1.

The HXD PIN instrument aboard \suzaku\ offers for the first time the
opportunity to investigate the variability of \3c\ at harder X-ray
energies.  Using the latest background file, we extracted light curves
in the 15--40 keV and 40--70 keV ranges; no variability was detected
in the harder energy band ($P_{\chi^2}=0.47$), while significant
variability ($P_{\chi^2}=1\times 10^{-4}$) seems to be present in the
15--40 keV energy band, which is shown in the bottom panel of
Fig.~\ref{figure:fig2}b.  However, if we increase the PIN background
level by applying systematic corrections at 1, 2 and 3 $\sigma$
levels, we obtain $P_{\chi^2}$ values of $1.6\times 10^{-2}$, 0.16,
and 0.52, respectively.  \suzaku\ observations thus indicate that
there is significant variability in the energy range 0.5--10 keV,
while above 15 keV the uncertainties on the HXD background prevent us
from drawing strong conclusions.

In summary, on short timescales (few hours) the flux of \3c\ is
consistent with the hypothesis of being constant in all energy
bands. On longer timescales (i.e., considering the two \xmm\
observations together or the 2-day \rxte\ and \suzaku\ coverage) the
flux variability becomes significant.
     
\section{Short-Time Scale Spectral Variability}
\begin{figure*} 
\includegraphics[bb=40 30 355 300,clip=,angle=0,width=6.cm]{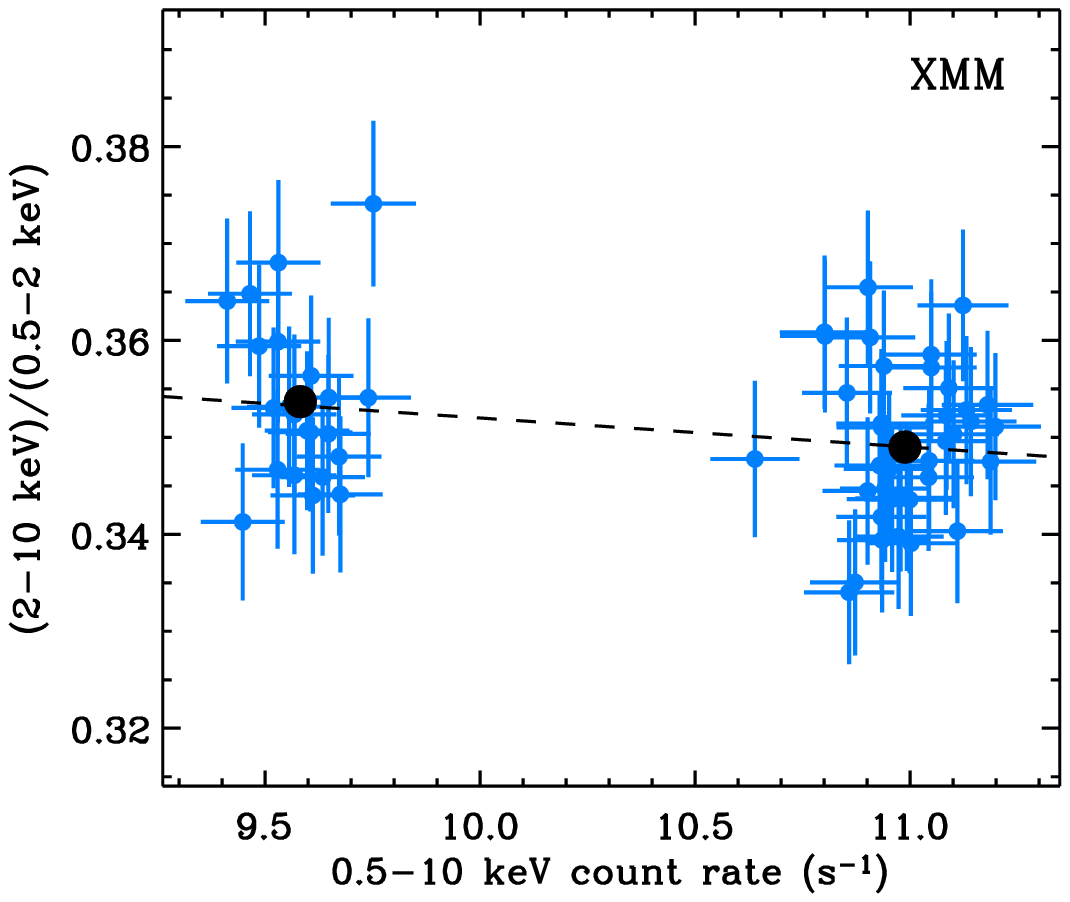}\includegraphics[bb=40 30 355 300,clip=,angle=0,width=6.cm]{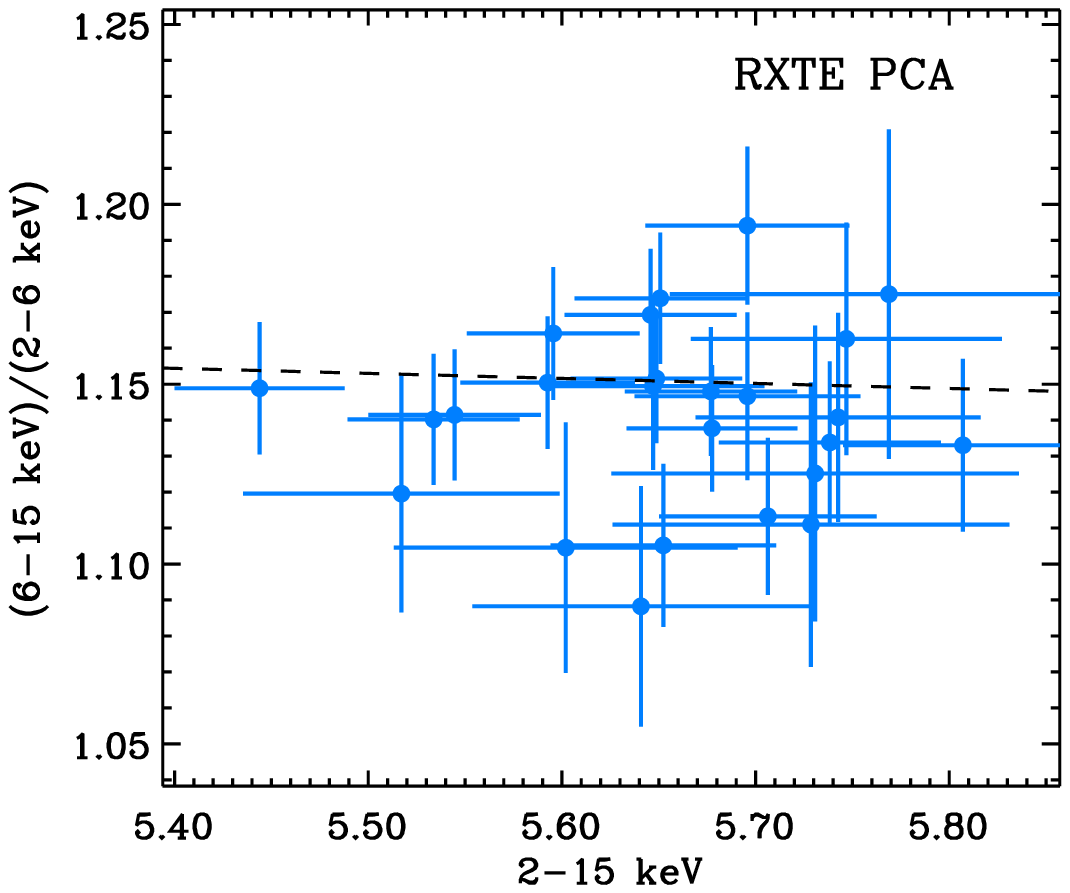}\includegraphics[bb=40 30 355 300,clip=,angle=0,width=6.cm]{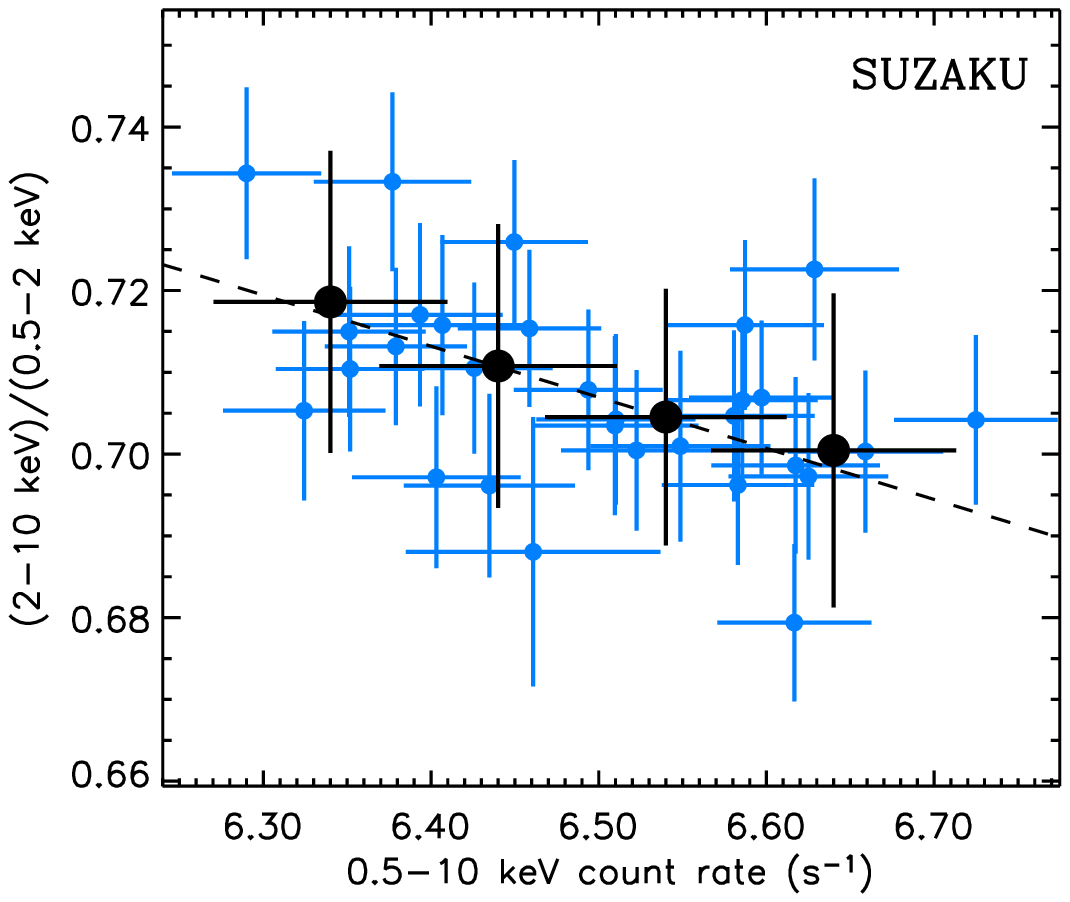} 
\caption{{\it Left:} \xmm\ hardness ratio (2--10 keV)/0.5--2 keV)
plotted versus the total count rate. The large black circles indicate the
average values during the 2 observations; the error bar is smaller than the 
symbols.
 The dashed line indicates the best 
linear fit. {\it Middle:} \rxte\ hardness ratio (6--15 keV)/2--6 keV)
plotted versus the total count rate. The dashed line indicates the best 
linear fit. 
{\it Right:} \suzaku\ XIS03 hardness ratio (2--10 keV)/0.5--2 keV)
plotted versus the total count rate. To guide the eye, binned values
(large black circles) have been plotted over non-binned values. 
The dashed line indicates the best 
linear fit:
$y=(1.11\pm0.13)-(0.06\pm0.02)x$.
} 
\label{figure:fig3} 
\end{figure*} 
In order to study the spectral variability of \3c, we use simple
methods such as the variation of the hardness ratio with time and with
total count rate, and the fractional variability in different energy
bands.  These can provide useful information without any a priori
assumption about the shape of the X-ray continuum. Thus, the results
from the study of these plots can be considered as
``model-independent".

To investigate the presence of spectral variability, we first apply a
$\chi^2$ test to the time series of the hardness ratio $HR=h/s$, where
$h=$2--10 keV and $s=$0.5--2 keV for \xmm\ and \suzaku, whereas for
\rxte\ $h=$6--15 keV and $s=$2--6 keV.  The results of this test are
reported in Table 2 and suggest that there is no significant spectral
variability on timescales of a few hours, while on a time scale of 2
days (\suzaku\ observation) marginally significant spectral
variability seems to be present.  It is worth noting that even between
\xmm\ observations A and B only marginally significant spectral
variability seems to occur: the average $HR$ increases from $0.349\pm
0.001$ to $0.354\pm 0.002$, which is a $2\sigma$ effect. This is a
direct consequence of the fact that the soft and hard count rates vary
by the same amount: they both decrease by $\sim$12\% in one week (see
Fig.~\ref{figure:fig1}).

Spectral variability can be further investigated by plotting $HR$ versus
the total count rate. Figure~\ref{figure:fig3} shows the $HR$-count
rate plots for \xmm\ (left panel), \rxte\ (middle panel), and \suzaku\
(right panel), respectively. The superimposed dashed lines represent
the best-fitting straight lines, which were obtained using the routine
\verb+fitexy+ (Press et al. 1997) that accounts for the errors not
only on the y-axis but along the x-axis as well.  Specifically, this
analysis yielded: $y=(0.38\pm0.02)-(0.003\pm0.002)x$ for \xmm,
$y=(1.23\pm0.41)-(0.01\pm0.07)x$ for \rxte\, and
$y=(1.11\pm0.13)-(0.06\pm0.02)x$ for \suzaku, respectively. In summary,
while \xmm\ and \rxte\ observations show very flat negative slopes
that are consistent with the hypothesis of constancy, \suzaku\ shows a
negative trend (i.e., a steepening of the spectrum with increasing 
source flux) that is significant at a $3\sigma$ level. This result
is confirmed by a non-parametric Spearman test that yields a rank correlation
value of -0.44 and a corresponding chance probability of $1.6\times10^{-2}$.

Another simple way to quantify the spectral variability of \3c,
without considering the time ordering of the values in the light
curves, is based on the fractional variability parameter $F_{\rm
var}$. This is a commonly used measure of the intrinsic variability
amplitude relative to the mean count rate, corrected for the effect of
random errors, i.e., 
\begin{equation} F_{\rm
var}={(\sigma^2-\Delta^2)^{1/2}\over\langle r\rangle} 
\end{equation}
where $\sigma^2$ is the variance, $\langle r\rangle$ the unweighted
mean count rate, and $\Delta^2$ the mean square value of the
uncertainty associated with each individual count rate. The error on
$F_{\rm var}$, reported in Table 2, has been estimated following
Vaughan et al. (2003). For all individual observations, we computed 
$F_{\rm var}$ on the soft and hard
energy bands (as defined above), since the relatively short
observations and the moderately low count rate do not allow this kind
of analysis on multiple narrow energy bands. The results, summarized
in Table 2, indicate that the variability amplitudes measured in the 2
energy bands are consistent with each other within the
uncertainties. During the \suzaku\ observation there is marginal
evidence ($\sim2.2\sigma$) that the soft band is more variable than
the hard one.

\section{Power Spectral Density Analysis}

To estimate the power-spectrum of the source in the 2--10 keV energy
band we used: 1) the 1999 and 2000 \rxte\ light curve, using a 3--day
bin size (``rxte-long" light curve, hereafter), 2) the 1996, 2-month
long, \rxte\ light curve, using a 1--day bin size (``rxte-medium"
light curve, hereafter), 3) the 2005, 2--day long, \rxte\ light curve
(``rxte-short" light curve, hereafter), 4) the 2006, 2--day long,
\suzaku\ XIS light curve, and 5) the October 10 and 17, 2004, \xmm\
EPIC pn light curves. We used a 5760 s binning for the rxte-short and
XIS light curves, and a 200 s binning for the \xmm\ light curves. The
October 10 2004, \xmm\ light curve was split in two parts to exclude
the $\sim 2$ hours period of enhanced background activity, which
was detected $\sim 7$ hours after the start of the observation.

All light curves are evenly sampled, with a few missing points (about
$5-10$\% of the total number of points). These missing points
are randomly distributed over each light curve, and we accounted for
them using a linear interpolation between the two bins adjacent to the
gaps, adding the appropriate Poisson noise in each case.

We used equation (1) in Papadakis \& Lawrence (1993) to compute the
periodograms of each light curve, after normalizing them to their
mean. The expected Poisson noise power level for the
rxte-long and rxte-medium light curves is comparable, and for this
reason we combined their periodograms in one file, and we sorted them
in order of decreasing frequency (the ``low-frequency"
periodogram). This combined periodogram can be used for the estimation
of the long and medium time scale power spectrum, from a frequency
$\sim 1$/(few days), down to $\sim 1$/(2 years). The rxte-short and
XIS light curves also have comparable Poisson noise power level, and
we therefore combined their periodograms forming the
``medium-frequency" periodogram to estimate the power spectrum at
higher frequencies of the order of $\sim 1/$(a few hours).  Finally,
we also combined the \xmm\ periodograms, in an attempt to detect a
source signal at even higher frequencies (the ``high frequency"
periodogram).

\begin{figure}
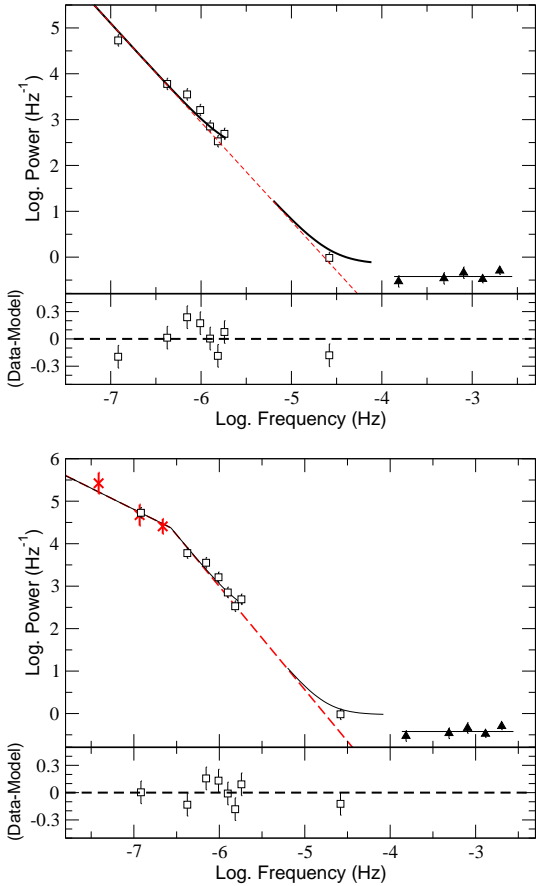
 
\includegraphics[bb=30 370 510 770,clip=,angle=0,height=6.cm]{f4a.eps}\hskip 1 cm\includegraphics[bb=30 370 510 770,clip=,angle=0,height=6.cm]{f4b.eps}
\caption{Power density spectra of \3c based on \rxte, \suzaku\ (open squares), 
and \xmm\ light curves (filled triangles) described in the text. 
The dashed lines indicate the best power law model fit (left
panel) and the best broken power-law model fit (right panel) to the full
band PSD. Solid lines indicate the best fit models when the Poisson
noise power level is also taken into account. Clearly, the intrinsic
power level is very low above $10^{-4}$ Hz, hence the observed PSD in the
XMM band is flat, and consistent with the predictions of a purely
Poisson noise power spectrum (indicated by the solid line at the highest
frequency part of the PSD).}
\label{figure:fig4} 
\end{figure}

Following  Papadakis \& Lawrence (1993),
we binned the three periodograms in log-log space, using bins
of size 20, and their equations (18), (19), and (20) (for the
estimation of the error of the resulting binned PSD points).
 Our results are plotted in Fig.~\ref{figure:fig4}. Filled
triangles indicate the high frequency PSD, estimated using the
high-frequency, \xmm\ periodogram. The solid line indicates the
expected Poisson power level. Clearly, at frequencies higher than
$10^{-4}$ Hz, we cannot detect any intrinsic variations. The open
squares in the same figure indicate the low and medium frequency PSD
estimates. At low frequencies, the source PSD shows the familiar
``red-noise" power spectral shape.

We fitted the low frequency PSD with a simple power-law model of the
form $P(f)\propto f^{-a}$, taking into account the different Poisson
noise power levels for the medium-frequency PSD estimate (indicated by
the open square around $f\sim 10^{-4.5}$ Hz) and the low-frequency PSD
(open squares at frequencies lower than $f\sim 10^{-5.7}$ Hz). 
We have also accounted for
aliasing effects, as they can be estimated analytically for any given
PSD model shape, following the analytical expressions in section 7.1.1
of Priestley (1989).
The model describes reasonably well the PSD: the best-fit slope is
$a=2.2\pm 0.2$ (errors correspond to 68\% confidence limits for two
interesting parameters), $\chi^2=12.8/6$ degrees of freedom (dof),
probability of null hypothesis, P$_{\rm null}$, of 4.6\%. The dashed
line in the left panel of Fig.~\ref{figure:fig4} indicates the
best-fit power-law model, and the thick solid black line, the best-fit
model after taking into account the different Poisson noise power
levels for the low and medium-frequency power spectra (best-fit
residuals are shown in the bottom panel of the same figure). The
best-fit slope is consistent with the high-frequency PSD slope
detected in radio-quiet Seyfert galaxies. The steepness of the
intrinsic power spectrum can explain the lack of detection of
intrinsic source variations in the \xmm\ light curves: the expected
amplitude at the highest frequencies we can probe is much smaller
than the amplitude caused by noise in the \xmm\ light curves.

We also tried to fit the PSD with a broken power-law model of the
form: $P(f)\propto (f/f_{\rm br})^{-b}$, where $f<f_{\rm br}$ and
$P(f)\propto (f/f_{\rm br})^{-a}$, at higher frequencies; $f_{\rm br}$
is the so called ``break-frequency". Such a model provides a good fit
to the PSD of many Seyfert galaxies with $a\sim 2$ and $b\sim 1$ (see,
e.g., Uttley, McHardy \& Papadakis 2002; Papadakis et al. 2002a;
Markowitz et al. 2003). Since the \3c\ PSD does not show a clear slope
change at low frequencies, we kept $b$ fixed at 1, which is the typical
value found in Seyfert galaxies. Note that, in case  $b$ is left 
free to vary,
all parameters are very poorly constrained, due to paucity of data
points at low frequencies. The best-fit
parameters are as follows: $a=2.4\pm0.3, f_{\rm
br}=2.7^{+3.6}_{-1.2}\times 10^{-7}$ Hz (68\% errors for 3 interesting
parameters: high frequency slope, break frequency, and normalization), 
$\chi^2=6.6/5$ dof, P$_{\rm null}$=25.2\%. The resulting
best-fit model is shown in the right panel of Fig.~\ref{figure:fig4}
(the symbols are the same as in the left panel), and indicates that it
fits the PSD very well. According to an F-test ($F_{\rm stat}=4.7$,
$P_F=0.082$), the broken power-law model provides an improvement of
the goodness of fit at a confidence level of  92\%,
compared to the power-law model. 
At the same time, the PSD plotted in the
right panel of Fig.~4 indicates that the detection of this
break-frequency in the power spectrum of the source is determined mainly
by the lowest frequency point in the PSD. The points indicated by the
crosses in the same panel, show the lowest frequency part of the power
spectrum. They correspond to the 20 lowest frequency periodogram
ordinates, and were estimated using bins of size 5 and 10, for the
higher frequency point. These points indicate that the low frequency end
of the PSD does follow a slope which is flatter than the slope of the
high frequency PSD. Nevertheless, to be
conservative, we consider the detection of a break-frequency in the
PSD of \3c\ as tentative.

\section{Discussion}

In order to put our results in perspective and better understand their
implications, it is important to have in mind the values of the
fundamental parameters that characterize the accretion process in \3c,
namely the black hole mass, $M_{\rm BH}$, and the accretion rate in
Eddington units, $\dot m$. To be consistent with the companion paper
from Sambruna et al. (2009), in the following we assume $M_{\rm
BH}=(5\pm1)\times 10^8~M_\odot$, which is based on the velocity
dispersion presented in Nelson et al. (2004) and is reported by  Lewis 
\& Eracleous (2006). However,  for completeness and given the large 
uncertainties, we 
also consider the value obtained via reverberation mapping 
$M_{\rm BH}=(2.87\pm0.64)\times 10^8~M_\odot$
(Peterson et al. 2004)
Once $M_{\rm BH}$ is determined, $\dot m$ readily follows, by
determining the bolometric luminosity, $L_{\rm bol}$, and dividing it
by the Eddington luminosity, $L_{\rm Edd}=1.3\times10^{38}(M_{\rm
BH}/M_\odot)$ \lum.  $L_{\rm bol}$ can be obtained by integrating the
broadband spectral energy distribution (SED) of \3c. A detailed
compilation of broadband data of \3c\ ranging from the radio to the
hard X-rays has been presented by Sambruna et al. (2009), yielding a
bolometric luminosity in the range 1--4 $\times 10^{45}$ \lum\
(this luminosity range is a direct consequence of the long-term
intrinsic variability of \3c). Combining these values with $M_{\rm
BH}$, we obtain an Eddington ratio ranging between 0.01 and 0.07
(0.03--0.1 for $M_{\rm BH}=2.87\times 10^8~M_{\odot}$), which
is consistent with the lower end of the accretion rate values
typically inferred in Seyfert 1 galaxies (see Vasudevan \& Fabian 2009
for a recent compilation). 
 For a source with powerful radio jets as \3c, it is
also important to know the inclination angle, since beamed emission
can affect any energy band including the X-rays. Based on radio data,
the jet inclination angle in \3c\ lies between 30\degr\ and 35\degr\ and
the estimated jet velocity is $\beta=v/c >0.96$  (Giovannini et al. 2001).
This translates into bulk Lorentz factors in the range $\Gamma=3.6-7.1$, 
which in turn yield beaming factors in the range $\delta\simeq 0.7-1.7$, 
implying that the jet emission can be either de-beamed or beamed but
in a moderate way.

The main aim of this work is to utilize the results from the
short-term variability and power spectral density analyses to derive
model-independent constraints on the central engine of \3c\ and
possibly on radio-loud AGN in general.  More specifically, our
findings can be used to evaluate the competing models proposed to
explain the weaker X-ray reprocessing features in BLRGs, namely
dilution from the base of the jet, radiatively inefficient flow (RIAF)
in the central region, or the presence of a highly ionized reflector.

\subsection{Flux variability}

Previous X-ray monitoring campaigns have shown that \3c\ is highly
variable on timescales ranging from several days to months.  For
example, using a 9-month light curve of \3c\ from the \rosat\ HRI
(which probes the 0.1--2.4 keV energy band), Leighly \& O'Brien (1997)
demonstrated that the soft X-rays are highly variable, displaying
large flares with quiescent periods in between, a behavior that and
implies a non-linear nature for the variability process. Similarly,
two long-term monitoring campaigns with \rxte, spanning respectively 3
months and 2 years, revealed that the source is highly variable also
at higher X-ray energies with $F_{\rm var}$ ranging
between 20 and 30\% (Gliozzi et al. 2003a; 2006).
However, since both \rosat\ and \rxte\ are low-Earth orbit satellites
the light curves were continuously interrupted, hampering the study of
variability on timescales of hours.

\xmm, with its highly elliptical orbit, has allowed the study of the
short-term variability of \3c\ for the first time. The EPIC light
curves showed a lack of any significant short-term variability,
despite the fact that the source was caught in a fairly high flux
state.
The lack of short-term variability lends support to the scenario where the
jet does not play an important role in the X-ray emission of \3c\
from 2 different points of view. First, it confirms on firmer statistical 
ground that \3c\ follows the scaling relations typical of radio-quiet AGN 
that predict the short-term variability to be negligible in AGN with 
large $M_{\rm BH}$. Second, it reveals a marked difference with respect
to the typical behavior observed in jet-dominated sources, which show strong
variability down to timescales of few minutes (e.g., Cui 2004). According
to the current blazar paradigm, the observed short-term variability at high
energies is a direct consequence of the small size of the emitting region
and the observed variability timescales are further shortened by beaming
 effects:
$t_{\rm obs}=t_{\rm rest}/\delta$. As a consequence, the lack of short-term
variability in \3c\ suggests that the X-ray emitting region is extended 
and that
the beaming effects, if present, are negligible in this energy band.

The 2-day long observations carried out with \rxte\ and \suzaku\ confirm
that on timescales longer than a few hours, the X-ray emission of \3c\
varies significantly.
Additionally, the analysis of \suzaku\ HXD-PIN data suggests a
possible detection of low-amplitude flux changes up to 40 keV.
However, the current
uncertainties on the HXD background prevent us from drawing strong
conclusions.

Important results from the temporal study can also be derived by the
PSD analysis, which is the best developed timing technique and the one 
commonly used
for investigating the time variability properties of Galactic black
hole systems (GBHs) and AGN. Indeed, recent detailed PSD studies have
been used to strengthen the link between GBHs and AGN (e.g., Uttley et
al. 2002; Markowitz et al. 2003, McHardy et al. 2004).

The comparison of the PSD frequency break $f_{\rm br}$ (indicative of
a characteristic timescale of the BH system) in GBHs and AGN has
provided an alternative way to determine the BH mass in AGN. McHardy
et al. (2006) proposed that, in Seyfert galaxies, the break timescale,
$T_{\rm br}=1/f_{\rm br}$, scales with $M_{\rm BH}$ and $L_{\rm bol}$,
following the relationship: $\log(T_{\rm br})=2.1\log(M_{\rm
BH})-0.98\log(L_{\rm bol})-2.32$, where $M_{\rm BH}$ is measured in
units of $10^6~M_{\odot}$, and $L_{\rm bol}$ in units of $10^{44}$
\lum. Using in the above formula 
$M_{\rm BH}=500\times 10^6~M_{\odot}$ and $L_{\rm bol}=(10-40)\times
10^{44}$ \lum, obtained from the direct integration of the SED, we
obtain that $T_{br}$ ranges between 60 and 234 days
(20--66 days for $M_{\rm BH}=287\times 10^6~M_{\odot}$).  Interestingly,
this result is consistent with the temporal break inferred from
our PSD analysis: $T_{br}=43^{+34}_{-25}$ days.

McHardy and collaborators (2006) also found a tight correlation 
between $T_{br}$ and the full width at half maximum (FWHM) of the $H_\beta$
line: $\log(T_{br})=4.20^{+0.71}_{-0.56} \times 
\log({\rm FWHM(H_\beta)})-14.43$. Taking into accout the uncertainties on the
slope, and using the value derived from our PSD analysis, $T_{br}=43$ d, the 
predicted FWHM($H_\beta$) for \3c\ ranges between $\sim$~2000 and 26,000
${\rm km~s^{-1}}$, which is fully consistent with the value of 12,700
${\rm km~s^{-1}}$ derived from a time-averaged spectrum over several years
(Sergeev et al. 2002).

The agreement between the predicted and the measured value of $T_{br}$
is important in two respects. First, it lends further support to the
detection of the break in the PSD of \3c. We do note that,
owing to the relatively short monitoring baseline of \3c\ (2 years as
opposed to 5--10 years used in Seyfert studies), the detection of a
break is significant only at the 92\% confidence level.
Nevertheless, we emphasize that this is the best estimate afforded by 
the current data and no improvement on the break determination will be 
possible in the years to come, since \3c\ is not monitored by any
X-ray satellite.  Second, and
perhaps more important, it suggests that the flux variability
properties of this BLRG are indistinguishable from those of
radio-quiet AGN.  In contrast, PSD studies of the most prominent
blazars, Mrk~421, Mrk~501, and PKS~$2155-304$ (i.e., jet-dominated AGN
that have been observed with \rxte\ in long monitoring campaigns)
suggest the presence of PSD breaks at frequencies that are nearly two
orders of magnitude higher than the tentative break found in \3c, i.e.,
at $f_{\rm br,blazar}\simeq10^{-5}$ Hz, or $T_{\rm br,blazar}<1$ day
(Kataoka et al. 2001). As a consequence,
the variability properties of \3c\ appear to be incompatible with jet
emission. On the other hand, the similarity between the variability
properties of Seyfert galaxies and \3c\ PSDs suggests that the X-ray
variability process, and by extension the X-ray emission mechanism are
similar between these two classes of object.

\subsection{Spectral Variability}

Previous studies, based on \rxte\ monitoring campaigns over periods
ranging from a few months to two years, have revealed that \3c\
shows correlated flux and spectral variations: the source spectrum
softens as the source becomes brighter (Gliozzi et al. 2003a, 2006),
which is the typical behavior observed in Seyfert galaxies (e.g.,
Papadakis et al. 2002b). Since these studies were performed in the
2--15 keV energy range, only our December 2005 \rxte\ observation can
be formally compared with previous results. Unfortunately, due to the
short duration of that observation and the limited range of the
observed flux variations (compared to the long monitoring campaigns),
no significant spectral variability is detected.

Thanks to the combination of XIS0, XIS1, and XIS3 data, and their
relatively low background level, the \suzaku\ light curves have higher
S/N than \rxte\ and the $HR$-count rate plot clearly indicates that \3c\ is
consistent with the typical Seyfert-like behavior over a time interval of 2
days. It is worth noting that, unlike \rxte\ that covers
only the hard X-ray range, \suzaku\ makes it possible to probe simultaneously
soft (i.e., E $<$ 2 keV) and hard energies. This is of crucial importance
for BLRGs that have generally complex X-ray spectra, which cannot be 
fitted with a simple power law  suggesting possible contributions from 
different physical components (see, e.g., Sambruna et al. 2009). For example,
based on spectral variability results, Kataoka et al. (2007) proposed that
the soft X-ray emission of 3C120 (another archetypal BLRG) was dominated by a 
jet. 

In this context, the fact that the \suzaku\ $HR-ct$ results 
are in full agreement with those obtained from an analogous analysis 
of long-term 2--15 keV \rxte\ data, suggests that also
at softer energies the emission is dominated by the same Comptonized 
component as in Seyfert galaxies. This conclusion is further confirmed by the
fact that soft (0.5--2 keV) and hard (2--10 keV) count rates appear to vary
in concert (see Fig. 2 right panel). This result does not exclude the possible
presence of an additional component with constant flux (such as some 
contribution from reflection), but it rules out the presence of a variable 
component, such as the beamed emission from a jet.

Further support for the latter conclusion comes from the spectral variability
analysis of \xmm\ data: soft and hard count rate appears to vary roughly
in concert not only within each single \xmm\ observation but also between the
2 \xmm\ pointings that are separated in time by 1 week and in flux by
12\%. At this point, one might wonder why \xmm\ with its superior throughput 
is unable to detect a statistical significant anti-correlation in the  $HR-ct$
plot at a confidence level higher than $2\sigma$. The reason is simply that on 
short timescales (individual exposures) count rate and $HR$ are 
constant, therefore all the information from one exposure virtually 
collapses to a single data point. As a consequence, the search for a negative
trend in the $HR-ct$ plot is based on 2 data points solely.

\section{Summary and Conclusions}

We have studied the short-term temporal and spectral variability
properties of the BLRG \3c\ using \xmm, \rxte, and \suzaku\
observations carried out between October 2004 and December 2006. Our
new data were then combined with older \rxte\ data obtained from
long-term monitoring campaigns to investigate the PSD in detail. The
main findings of our analysis can be summarized as follows:

\begin{itemize}

\item 
On short timescales (i.e., few hours, probed by uninterrupted \xmm\
observations) the flux of \3c\ in all energy bands is consistent with
the hypothesis of being constant. On longer timescales (i.e.,
considering the two \xmm\ observations together or the 2-day \rxte\
and \suzaku\ coverage) the flux variability becomes significant.

\item
A detailed PSD analysis carried out over five decades in frequency
suggests the presence of a break at $T_{br}=43^{+34}_{-25}$ days at a
92\% confidence level. This is the second tentative detection of a PSD
break in a radio-loud, non-jet dominated AGN, after the recent results
of the BLRG 3C~120 from Marshall et al. (2009). Importantly, the time
scale corresponding to the break frequency is in agreement with the
relation between $T_{br}$, $M_{\rm BH}$, and $L_{\rm bol}$
as well as with the $T_{br}-{\rm FWHM(H_\beta)}$ relationship, 
both of which are valid for
Seyfert galaxies. Note that, while the relative brevity of the long-term
\rxte\ campaign hampers the significance of the break detection around
40 days, the quality of the data is sufficient to rule out
the presence of a break at shorter timescales, which is typically
detected in PSDs of jet-dominated sources.

\item
The 2-day long \suzaku\ observation indicates that \3c\ shows the
typical spectral evolution of Seyfert galaxies in the 0.5--10 keV
energy range (the spectrum becomes softer as the source brightens).
This confirms previous results, based on long \rxte\ monitoring 
campaigns, and expands them to different energy bands and to much
shorter timescales.

\item
The broadly coordinated variability in soft and hard X-rays 
during the 2-day \suzaku\ observation and between
the 2 \xmm\ pointings, taken one week apart, suggests a common
physical origin for both energy bands, arguing against the presence of
an additional variable component (i.e., a jet) emerging at softer
energies.

\item
The lack of short-term flux variability, the frequency break of the
PSD, and the Seyfert-like spectral variability consistently argue
against a scenario where a jet plays a significant role in the X-ray
regime, confirming the results from the time-averaged spectral
analysis (Sambruna et al. 2009).

\end{itemize}

In conclusion, all our results indicate that the flux variability
properties of \3c\ are broadly consistent with those of radio-quiet
AGN, suggesting that the X-ray variability process and, by extension,
the emission mechanism in \3c\ is similar to that of Seyfert galaxies.
This, in turn, suggests that the weaker reflection features observed
in the X-ray spectrum of 3C~390.3 are not a result of dilution by
jet emission, in agreement with the conclusions of the spectral
analysis of Sambruna et al. (2009). If variability studies of other
BLRGs show similar results, then jet dilution will be disfavored as a
general explanation of the weak reflection features in all BLRGs as a
class.  Of course, the jet can still influence the observed X-ray
properties of BLRGs by obscuring the central regions of the accretion
disk or by beaming radiation away from the disk surface. The
obscuration scenario has been suggested by Sambruna et al. (2009) and
Larsson et al. (2008) to account for the lack of reflection from the
inner accretion disk in 3C~390.3 and 4C+74.26,
respectively. Unfortunately, this scenario would also block from view
the region of the accretion flow where the jet is formed. Significant
progress can be made by deep broadband observations of a large number
of BLRGs so as to sample a wide variety of accretion disk
geometries. The variability analysis presented here shows that, as
long as the jet angle to line of sight is large enough, such an
investigation will not be subject to the effects of jet dilution.



\begin{thebibliography}{}

\bibitem[Ballatyne et al. 2004]{ball04} Ballantyne, D.R., Fabian, A.C.,
\& Iwasawa, K. 2004, MNRAS, 354, 839

\bibitem[Ballatyne 2005]{ball05} Ballantyne, D.R. MNRAS, 362, 1183

\bibitem[Bennet et al. 2003]{ben03} Bennet, C.L. et al. 2003, ApJS, 148, 1


\bibitem[Cui 2004]{cui04} Cui, W. 2004, ApJ, 605, 662

\bibitem[Eracleous et al. 1996]{eracl00} Eracleous, M., Halpern, J.P., 
\& Livio, M. 1996, ApJ, 459, 89

\bibitem[Eracleous et al. 2000]{eracl00} Eracleous, M., Sambrunna, R., 
\& Mushotzky, R.F. 2000, ApJ, 537, 654


\bibitem[Giovannini et al. 2001]{giova01} Giovannini, G., Cotton, W.D.,
Feretti, L., Lara, L., \& Venturi, T. 2001, ApJ, 552, 508


\bibitem[Gliozzi et al. 2003a]{glio03a} Gliozzi, M., Sambruna, R.M., \& 
Eracleous, M. 2003a, ApJ, 584, 176

\bibitem[Gliozzi et al. 2003b]{glio03b} Gliozzi, M., Sambruna, R.M., \& 
Brandt, W.N. 2003b, A\&A, 408, 949


\bibitem[Gliozzi et al. 2006]{glio06} Gliozzi, M., Papadakis, I.E., \& Raeth, 
C. 2006, A\&A, 449, 969

\bibitem[Gliozzi et al. 2007]{glio07} Gliozzi, M., et al. 2007, ApJ, 664, 88

\bibitem[Gliozzi et al. 2008]{glio08} Gliozzi, M., Foschini, L.,
Sambruna, R.M., \& Tavecchio, F. 2008, A\&A, 478, 723

\bibitem[Grandi et al. 1999]{grand99} Grandi, P., et al. 1999, A\&A, 343, 33

\bibitem[Grandi et al. 2006]{grand06} Grandi, P., Malaguti, G., \& Fiocchi, M. 2006, ApJ, 642, 113

\bibitem[Inda et al. 1994]{inda94} Inda, M., et al. 1994, ApJ, 420, 143

\bibitem[Jahoda et al. 1996]{jah96} Jahoda, K., Swank, J., Giles,
A.B., et al. 1996, Proc.SPIE, 2808, 59

\bibitem[Kataoka et al. 2001]{kata01} Kataoka, J., et al. 2001, ApJ, 560, 659

\bibitem[Kataoka et al. 2007]{kata07} Kataoka, J., et al. 2007, PASJ, 59, 279

\bibitem[Larsson et al. 2008]{lars08}  Larsson, J., Fabian, A.C., Ballantyne, 
D.R., \& Miniutti, G. 2008, MNRAS, 388, 1037

\bibitem[Leighly et al. 1997]{leigh97} Leighly, K.M., et al. 1997, ApJ, 483

\bibitem[Leighly \& O'Brien 1997]{leigh97b} Leighly, K.M. \& O'Brien, P.T. 1997, ApJ, 
481, L15 

\bibitem[Lewis et al. 2005]{lewis05} Lewis, K.T., et al. 2005, ApJ, 622, 816

\bibitem[Lewis \& Eracleous 2006]{lewis06} Lewis, K.T. \& Eracleous M. 2006, 
ApJ, 642, 711

\bibitem[Markowitz et al 2003]{markow03} Markowitz, A. et al. 2003, ApJ, 593,
96

\bibitem[Marshall et al. 2009]{marsh09} Marshall, K., et al. 2009, ApJ, 696, 
601

\bibitem[McHardy et al. 2004]{mcard04} McHardy, I., et al. 2004, MNRAS, 348, 783
\bibitem[McHardy et al. 2006]{mcard06} McHardy, I., et al. 2006, Nature, 444, 730

\bibitem[Nelson et al. 2004]{nelo04} Nelson, C.H., Green, R.F., Bower, G.,
Gebhardt, K., \& Weistrop, D. 2004, ApJ, 615, 652

\bibitem[Ogle et al. 2004]{ogle04} Ogle, P.M. et al. 2004, ApJ, 618, 139

\bibitem[1993]{papad1} Papadakis, I.E. \& Lawrence, A. 1993, MNRAS, 261, 612


\bibitem[Papadakis et al. 2002a]{papad02} Papadakis, I.E. et al. 2002a, A\&A, 
382, L1

\bibitem[Papadakis et al. 2002b]{papad02b} Papadakis, I.E. et al. 2002b, ApJ, 
573, 92

\bibitem[Papadakis et al. 2008]{papad08} Papadakis, I.E.,Ioannou, Z., 
Brinkmann, W., \& Xilouris, E.M. 2008, A\&A, 490, 995


\bibitem[Peterson et al. 2004]{pete04} Peterson, B.M., et al. 2004, ApJ, 613, 
682

\bibitem[Press et al. 1997]{press97} Press, W.H., Teukolsky, S.A., Vetterling, W.T., \& Flannery, B.P. 1997, Numerical Recipes (Cambridge: Cambridge Univ. Press)

\bibitem[Priestley 1989]{priest89}  Priestley, M.B. 1989, Spectral Analysis and Time
Series, Academic Press, London.

\bibitem[Rotschild et al. 1998]{rot98} Rotschild, R.E., Blanco, P.R., Gruber, D.E., et
al. 1998, ApJ, 496, 538

\bibitem[Sambruna et al. 1999]{samb99} Sambruna, R.M., Eracleous, M., \& Mushotzky, R. 1999, 
ApJ, 526, 60

\bibitem[Sergeev et al. 2002]{serg02} Sergeev, S.G., Pronik, V.I., Peterson, 
B.M., Sergeeva, E.A., \& Zheng, W. 2002, ApJ, 576, 660

\bibitem[Str\"uder et al. 2001]{str01} Str\"uder, L., Briel, U., Dennerl, K., 
et al. 2001, A\&A, 365, L18

\bibitem[Turner et al. 2001]{tur01} Turner, M.J., Abbey, A., Arnaud, M., et 
al. 2001, A\&A, 365, L27

\bibitem[Uttley et al.2002]{ut02} Uttley, P., McHardy, I., \& Papadakis, I.E.
2002, MNRAS, 332, 231

\bibitem[Vasudevan \& Fabian 2008]{vasu08} Vasudevan, R.V. \& Fabian, A.C. 
2008, MNRAS, 392, 1124

\bibitem[Vaughan et al. 2003]{vaugh03} Vaughan, S., Edelson, R., Warwick, R.S.,
\& Uttley, P. 2003, MNRAS, 345, 1271

\bibitem[Wo\'zniak et al. 1998]{woz98} Wo\'zniak, P.R., Zdziarski, A.A., Smith, D., Madejski, G.M., \& 
Johnson, W.N., 1998, MNRAS, 299, 449

\bibitem[Zdziarski \& Grandi 2001]{zdz01} Zdziarski, A.A. \& Grandi, P. 2001, ApJ, 551, 186

\end{thebibliography}
\end{document}